\documentclass[aps,twocolumn,showpacs]{revtex4}
\usepackage{graphicx}
\usepackage{amsmath}
\usepackage{amsfonts}
\usepackage{color}
\newcommand{\tbox}[1]{\mbox{\tiny #1}}

\begin{document}



\title{Tunable subdiffusion in the Caputo fractional standard map}


\author{J. A. M\'endez-Berm\'udez}
\address{Instituto de F\'isica, Benem\'erita Universidad Aut\'onoma de Puebla, 
Puebla 72570, Mexico}

\author{R. Aguilar-S\'anchez}
\address{Facultad de Ciencias Qu\'imicas, Benem\'erita Universidad Aut\'onoma de Puebla,
Puebla 72570, Mexico}

\begin{abstract}
The Caputo fractional standard map (C-fSM) is a two-dimensional nonlinear map with memory 
given in action-angle variables $(I,\theta)$.  
It is parameterized by $K$ and $\alpha\in(1,2]$ which control the strength of nonlinearity 
and the fractional order of the Caputo derivative, respectively. 
In this work we perform a scaling study of the average squared action $\left< I^2 \right>$ 
along strongly chaotic orbits, i.e.~when~$K\gg1$.
We numerically prove that $\left< I^2 \right>\propto n^\mu$ with $0\le\mu(\alpha)\le1$, for 
large enough discrete times $n$. 
That is, we demonstrate that the C-fSM displays subdiffusion for $1<\alpha<2$.
Specifically, we show that diffusion is suppressed for $\alpha\to1$ since  
$\mu(1)=0$, while standard diffusion is recovered for $\alpha=2$ where $\mu(2)=1$.
We describe our numerical results with a phenomenological analytical estimation.
We also contrast the C-fSM with the Riemann-Liouville fSM and Chirikov's standard map.
\end{abstract}

\maketitle

\section{Preliminaries}

By replacing the second order derivative in the equation of motion of the kicked rotor 
\begin{equation}
    \frac{d^2\theta}{dt^2} + K \sin(\theta) \sum_{j = 0}^\infty \delta\left(\frac{t}{T} - j \right) = 0
    \label{KR}
\end{equation}
by fractional
operators (fractional derivatives, fractional integrals or fractional integro-differential operators), 
fractional versions of the kicked rotor are obtained. 
The kicked rotor represents a free rotating stick in an inhomogeneous field that is periodically 
switched on in instantaneous pulses, see e.g.~\cite{O08}.
In Eq.~(\ref{KR}), $\theta\in [0,2\pi]$ is the angular position of the stick, $K$ is the kicking strength, 
$T$ is the kicking period, and $\delta$ is Dirac's delta function.
Among the several fractional kicked rotors (fKRs) reported in the literature we can mention:
the Riemann-Liouville fKR~\cite{TZ08,ET09}
\begin{equation}
_0D_t^\alpha \theta + K \sin(\theta) \sum_{j = 0}^\infty \delta\left(\frac{t}{T}  - (j+\epsilon) \right) = 0 , \quad 1< \alpha \leq 2 , 
\label{fKR}
\end{equation}
where $\epsilon\to0+$,
the Caputo fKR~\cite{T09,E11}
\begin{equation}
_0{^CD}_t^\alpha \theta + K \sin(\theta) \sum_{j = 0}^\infty \delta\left(\frac{t}{T}  -  (j+\epsilon) \right) = 0 , \quad 1< \alpha \leq 2 ,
\label{fKR}
\end{equation}
where $\epsilon\to0+$,
the Hadamard fKR~\cite{T21d}, 
the Erdelyi-Kober fKR~\cite{T21e}, and 
the Hilfer fKR~\cite{T21f}.
Above~\cite{SKM93,KST06},
\begin{align}
&_0D_t^\alpha \theta(t) =  D_t^m {_0 {\cal I}}_t^{m-\alpha}\theta(t) \nonumber \\ 
& = \frac{1}{\Gamma(m-\alpha)}\frac{d^{m}}{dt^{m}}\int_{0}^{t}\frac{\theta(\tau)d\tau}{(t-\tau)^{\alpha-m+1}}, \quad m-1<\alpha\leq m, \nonumber
\end{align}
\begin{align}
&_0{^CD}_t^\alpha \theta(t) = {_0 {\cal I}}_t^{m-\alpha}D_t^m\theta(t) \nonumber \\ 
& = \frac{1}{\Gamma(m-\alpha)}\int_{0}^{t}\frac{D_t^m\theta(\tau)d\tau}{(t-\tau)^{\alpha-m+1}}, \quad m-1<\alpha\leq m, \nonumber
\end{align}
with $D_t^m = d^m/dt^m$, $_0{\cal I}_t^m f(t)$ is a fractional integral given by
\begin{equation*}
_0{\cal I}_t^m f(t) = \frac{1}{\Gamma(m)}\int_0^t (t-\tau)^{\alpha-1}f(\tau)d\tau ,
\end{equation*}
and $\Gamma$ is the Gamma function.

All the fKRs listed above, have stroboscopic versions which are two-dimensional nonlinear 
maps with memory given in action-angle variables $(I,\theta)$. These maps are named as fractional 
standard maps (fSMs), in resemblance with Chirikov's standard map (CSM)~\cite{C69}:
\begin{equation}
\begin{array}{ll}
I_{n+1} = I_n - K\sin(\theta_n) , \\
\theta_{n+1} = \theta_n + I_{n+1}, \qquad \qquad \mbox{mod}(2\pi);
\end{array}
\label{CSM}
\end{equation}
which is the stroboscopic version of the standard kicked rotor of Eq.~(\ref{KR}).
Here and below, $T$ is set to one.

\begin{figure*}[ht]
\centering
\includegraphics[width=0.9\textwidth]{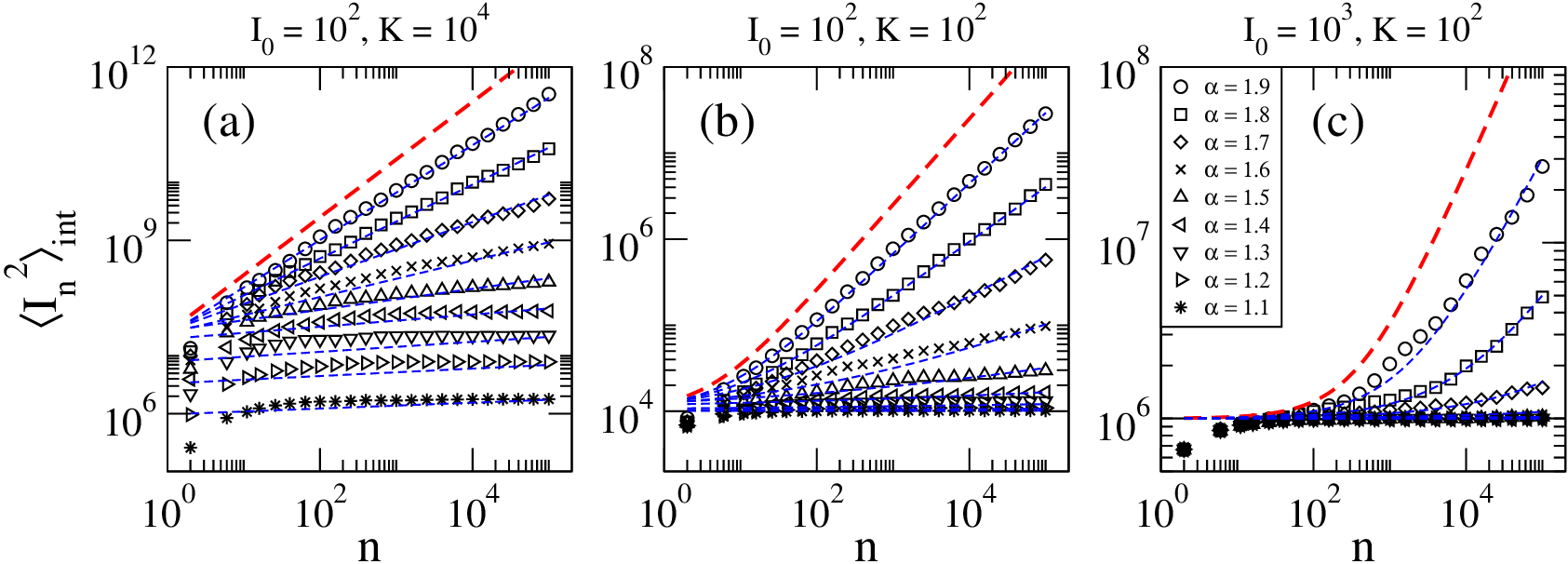}
\caption{Average squared action $\left< I^2_n \right>_{\tbox{int}}$ as a function of $n$ for 
(a) $(I_0,K)=(10^2,10^4)$, (b) $(I_0,K)=(10^2,10^2)$, and (c) $(I_0,K)=(10^3,10^2)$.
Several values of $\alpha$ are considered, as indicated in panel (c).
Red-dashed lines correspond to Eq.~(\ref{I2RLfSM}).
Blue-dashed lines are Eq.~(\ref{I2intCfSM}).
The average is taken over $M=200$ orbits with initial random phases in the interval $0<\theta_0<2\pi$.}
\label{Fig02}
\end{figure*}

As far as we know, the first two fSMs reported in the literature are the Riemann-Liouville fSM 
(RL-fSM)~\cite{TZ08,ET09},
\begin{equation}
\begin{array}{ll}
I_{n+1} = I_n - K\sin(\theta_n) , \\
\theta_{n+1} = \displaystyle{ \frac{1}{\Gamma(\alpha)}\sum_{i = 0}^{n} I_{i+1} V^1_\alpha(n-i+1)}, \quad \mbox{mod}(2\pi),
\end{array}
\label{RLfSM}
\end{equation}
and the Caputo fSM (C-fSM)~\cite{T09,E11},
\begin{equation}
\begin{array}{ll}
I_{n+1} = I_n \\ 
\quad \quad \ - \displaystyle \frac{K}{\Gamma(\alpha-1)} \left[ \sum_{i = 0}^{n-1} V^2_\alpha(n-i+1)\sin(\theta_i) + \sin(\theta_n) \right]  , \\
\theta_{n+1} = \theta_n + I_0 \\
\quad \quad \ - \displaystyle \frac{K}{\Gamma(\alpha)} \sum_{i = 0}^{n} V^1_\alpha(n-i+1)\sin(\theta_i), \quad \mbox{mod}(2\pi).
\end{array}
\label{CfSM}
\end{equation}
Here, $1<\alpha \leq 2$ is assumed and
\begin{equation*}
	V^k_\alpha(m) = m^{\alpha-k}-(m-1)^{\alpha-k}.
\end{equation*}
Both, the RL-fSM and the C-fSM are parameterized by $K$ and $\alpha$ which control the strength of 
nonlinearity and the fractional order of the derivative, respectively.
For $\alpha=2$, both the RL-fSM and the C-fSM reproduce the CSM~\cite{C69,E11}.

As compared with the CSM, which presents the generic transition to chaos (in the context of 
Kolmogorov--Arnold--Moser theorem, see e.g.~\cite{O08}), depending on the parameter pair ($K,\alpha$),
the RL-fSM and the C-fSM show richer dynamics: They generate attractors (fixed points, asymptotically 
stable periodic trajectories, slow converging and slow diverging trajectories, ballistic 
trajectories, and fractal-like structures) and/or chaotic trajectories~\cite{ET09,ET13,E19,E11}. 

Among several available studies on the RL-fSM and the C-fSM (see e.g.~\cite{ET09,ET13,E19,E11}), 
very recently, the squared average action $\left< I^2_n \right>$ of the RL-fSM was analyzed in the regime 
of $K\gg1$~\cite{MASL23}. There it was shown that, for strongly chaotic orbits, $\left< I^2_n \right>$ presents
normal diffusion (for sufficiently large times) and, in addition, it does not depend on $\alpha$. Indeed, the 
panorama reported for 
$\left< I^2_n \right>$ vs.~$n$ for the RL-fSM~\cite{MASL23} is equivalent to that of the CSM~\cite{MOL16,LS07} 
as well as that of the discontinuous standard map (DSM)~\cite{MOL16,MA12}, both with $K\gg 1$.
Moreover, an analytical estimation~\cite{MASL23}, used to get
\begin{equation}
\label{I2RLfSM}
\left< I^2_n \right>_{\tbox{RL-fSM}} = I_0^2 + \frac{K^2}{2} n ,
\end{equation}
also showed the independence of $\left< I^2_n \right>$ on $\alpha$.

By following the derivation of Eq.~(\ref{I2RLfSM}) we have realized that the independence of 
$\left< I^2_n \right>$ on $\alpha$ is due to the absence of $\alpha$ in the first equation of 
map~(\ref{RLfSM}). That is way Eq.~(\ref{I2RLfSM}) also describes the dynamics of CSM: note 
that the equation for the action is the same in both maps; see Eqs.~(\ref{CSM}) and~(\ref{RLfSM}).
This suggests that $\left< I^2_n \right>$ may depend on $\alpha$ in fractional maps 
where $\alpha$ appears in the equation for the action, such as map~(\ref{CfSM}).
Unfortunately, by the use of simple arguments as those used to get Eq.~(\ref{I2RLfSM}) in
Ref.~\cite{MASL23}, 
we are not able to get an explicit expression for $\left< I^2_n \right>$ for the C-fSM. 

Therefore, the purpose of this work is twofold. 
First, we numerically look for the effects of $\alpha$ on $\left< I^2_n \right>$ for the C-fSM,
$\left< I^2_n \right>_{\tbox{C-fSM}}$.
Second, we derive a phenomenological expression for $\left< I^2_n \right>_{\tbox{C-fSM}}$ 
which properly incorporates the parameter $\alpha$.

\begin{figure*}[ht]
\centering
\includegraphics[width=0.65\textwidth]{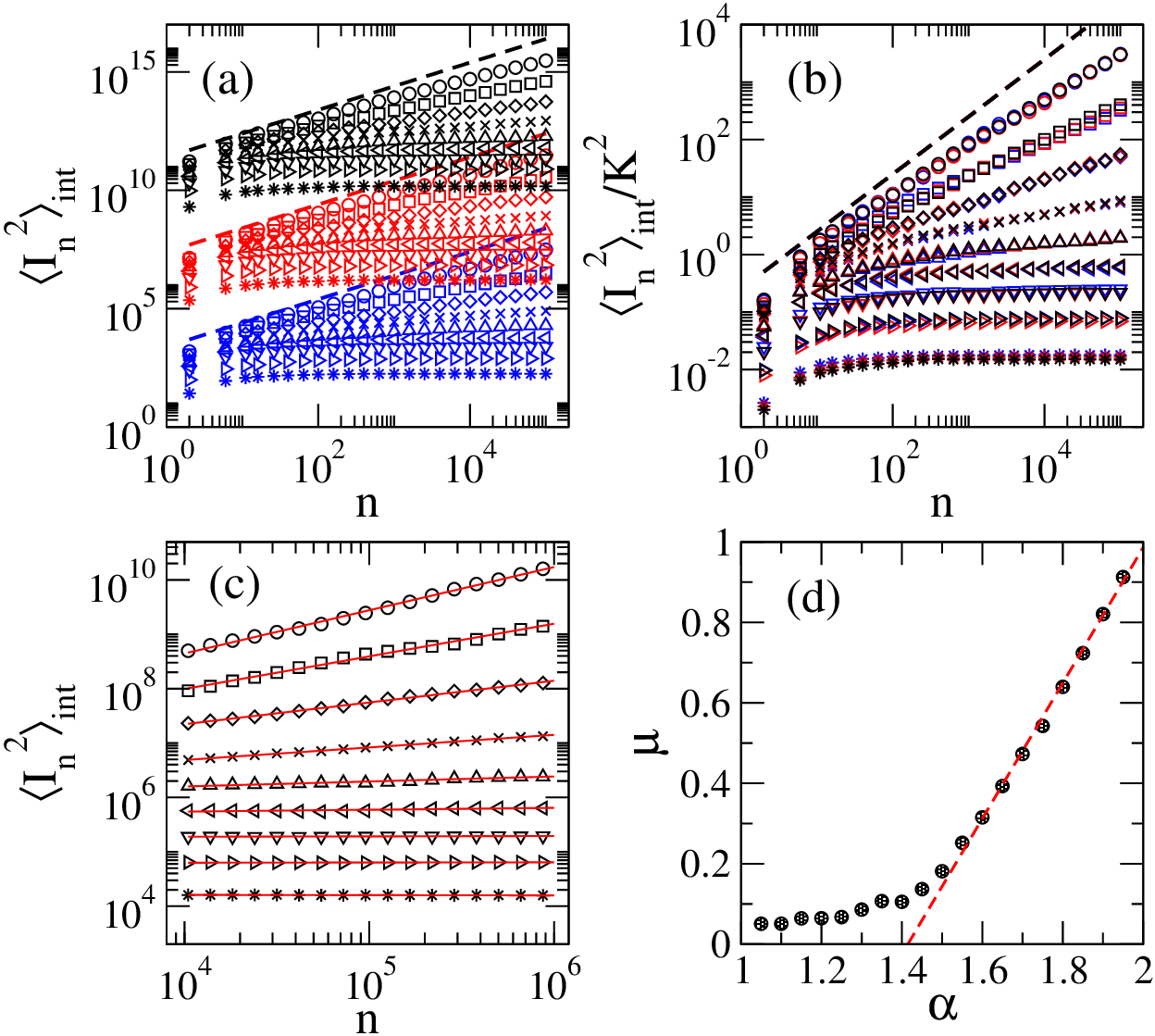}
\caption{(a) Average squared action $\left< I^2_n \right>_{\tbox{int}}$ as a function of $n$ 
for $K=10^2$ (blue symbols), $K=10^4$ (red symbols), and $K=10^6$ (black symbols). 
In all cases $I_0=0$.
The average is taken over $M=200$ orbits with initial random phases in the interval $0<\theta_0<2\pi$.
Several values of $\alpha$ are considered; same symbol labeling as in Fig.~\ref{Fig02}.
Dashed lines correspond to Eq.~(\ref{I2RLfSM}).
(b) $\left< I^2_n \right>_{\tbox{int}}/K^2$ vs.~$n$. Same data as in panel (a).
(c) $\left< I^2_n \right>_{\tbox{int}}$ vs.~$n$ for $K=10^3$ and $I_0=0$.
Here the average is taken over $M=100$ orbits with initial random phases in the interval $0<\theta_0<2\pi$.
Dashed lines correspond to power-law fittings of the form $\left< I^2_n \right>_{\tbox{int}}\propto n^\mu$
in the interval $n=[10^4,10^6]$.
(d) $\mu$, from the power-law fittings of panel (c), as a function of $\alpha$.
The red-dashed line is a linear fit to the data with $\alpha>0.5$: $\mu\sim1.69\alpha$.}
\label{Fig03}
\end{figure*}

\section{On the effects of $\alpha$ on $\left< I^2_n \right>_{\tbox{C-fSM}}$}

To ease our numerical analysis, to get curves smoother than the present 
$\left< I^2_n \right>_{\tbox{C-fSM}}$ vs.~$n$,
in what follows we compute the cumulative-normalized value of $\left< I^2_n \right>_{\tbox{C-fSM}}$,
\begin{equation*}
\left< I^2_n \right>_{\tbox{int}} = \frac{1}{n} \int^n_{n_0=0} \left< I^2_{n'} \right>_{\tbox{C-fSM}} dn' ,
\label{I2int}
\end{equation*}
by averaging over $M$ independent orbits (by randomly choosing values 
of $\theta_0$ in the interval $0<\theta_0<2\pi$) for each combination of parameters $(I_0,K,\alpha)$.

Then, in Fig.~\ref{Fig02} we plot $\left< I^2_n \right>_{\tbox{int}}$ as a function of $n$ for the C-fSM for 
several values of $\alpha$ in the interval $1<\alpha< 2$.
Moreover, in all panels we include Eq.~(\ref{I2RLfSM}) (as red-dashed curves) which corresponds
to the case $\alpha=2$; so we can contrast the results for the C-fSM with those for the RL-fSM~\cite{MASL23},
the CSM~\cite{MOL16,LS07}, and the DSM~\cite{MOL16,MA12}.
In Fig.~\ref{Fig02} we use three representative parameter pairs $(I_0,K)$: 
$I_0 < K$ (left panel), $I_0 = K$ (central panel), and $I_0 > K$ (right panel).

From Fig.~\ref{Fig02} we can clearly observe that $\alpha$ supressess the action diffusion even at the 
very first iteration; moreover, the smaller the value of $\alpha$ the larger the difference between 
$\left< I^2_n \right>_{\tbox{int}}$ and the red-dashed curves which correspond to normal diffusion. 
Also, for large iteration times $\left< I^2_n \right>_{\tbox{int}}$ grows proportional to 
$n^\mu$ with $\mu\equiv\mu(\alpha)$; this can be better observed in Fig.~\ref{Fig02}(a).
In addition, we observe two scenarios depending on the initial action $I_0$ as compared with $K$.
Specifically, when $I_0 < K$, the curves $\left< I^2_n \right>_{\tbox{int}}$ vs.~$n$ are all different for
different $\alpha$ and approach faster the regime $\left< I^2_n \right>_{\tbox{int}} \propto n^\mu$; see 
e.g.~Fig.~\ref{Fig02}(a).
While for $I_0 > K$, first, the curves $\left< I^2_n \right>_{\tbox{int}}$ vs.~$n$ for different $\alpha$ 
fall one on top of the other up to a crossover time $n^*$, after which $\left< I^2_n \right>_{\tbox{int}}$ 
grows proportional to $n^\mu$; see e.g.~Fig.~\ref{Fig02}(c).

In what follows we concentrate on the case $I_0 < K$ to easily approach the asymptotic regime 
where $\left< I^2_n \right>_{\tbox{int}} \propto n^\mu$.
So, in Fig.~\ref{Fig03}(a) we show $\left< I^2_n \right>_{\tbox{int}}$ as a function of $n$ for several
values of $\alpha$ and $I_0=0$.
Here we have used three values of $K$: $K=10^2$ (blue symbols), $K=10^4$ (red symbols), and 
$K=10^6$ (black symbols). 
Note that the contribution of $K$ to $\left< I^2_n \right>_{\tbox{int}}$ is through the factor $K^\gamma$, 
i.e.~$\left< I^2_n \right>_{\tbox{int}} \propto K^\gamma n^\mu$,
where $\gamma$ should be equal to 2, see e.g.~Eq.~(\ref{I2RLfSM}).
We verify this last statement in Fig.~\ref{Fig03}(b) where we plot the same curves of panel (a) but
now divided by $K^2$ and observe that curves for the same $\alpha$ fall one on top of the other.

Then, to characterize the dependence of $\mu$ on $\alpha$ in the asymptotic regime, i.e.~where 
$\left< I^2_n \right>_{\tbox{int}} \propto n^\mu$, in Fig.~\ref{Fig03}(c) we look at large iteration times.
There, we perform power-law fittings of the form $\left< I^2_n \right>_{\tbox{int}}\propto n^\mu$ in 
the interval $n=[10^4,10^6]$. The values of $\mu$ obtained from the fittings are reported in 
Fig.~\ref{Fig03}(d).
From Fig.~\ref{Fig03}(d) we can see that $\mu\to 0$ for $\alpha\to 1$ while $\mu\to 1$ for $\alpha\to 2$.
In addition we observe that $\mu(\alpha)\propto\alpha$ for $\alpha>0.5$.

Indeed, by substituting $\alpha=1$ into Eq.~(\ref{CfSM}), since $\Gamma(0)$ diverges the action
remains constant, $I_n=I_0$, so the action diffusion is fully suppressed and $\mu(\alpha=1)=0$. 
While substituting $\alpha=2$ into map~(\ref{CfSM}), since $\Gamma(1)=1$ 
and $V^2_2(m)=0$, the equation for to action reduces to $I_{n+1}=I_n-K\sin(\theta_n)$; so 
$\left< I^2_n \right>$ is described by Eq.~(\ref{I2RLfSM}) and $\mu(\alpha=2)=1$.
Therefore, for $1<\alpha<2$ the C-fSM shows subdiffusion: 
\begin{equation}
\left< I^2_n \right>_{\tbox{int}} \propto K^2n^{\mu(\alpha)} \quad \mbox{with} \quad 0<\mu(\alpha)<1 ,
\label{scalingA}
\end{equation}
which can be observed for large enough $n$.

\section{Heuristic estimate of $\left< I^2_n \right>_{\tbox{C-fSM}}$}

\begin{figure}[t!]
\centering
\includegraphics[width=0.8\columnwidth]{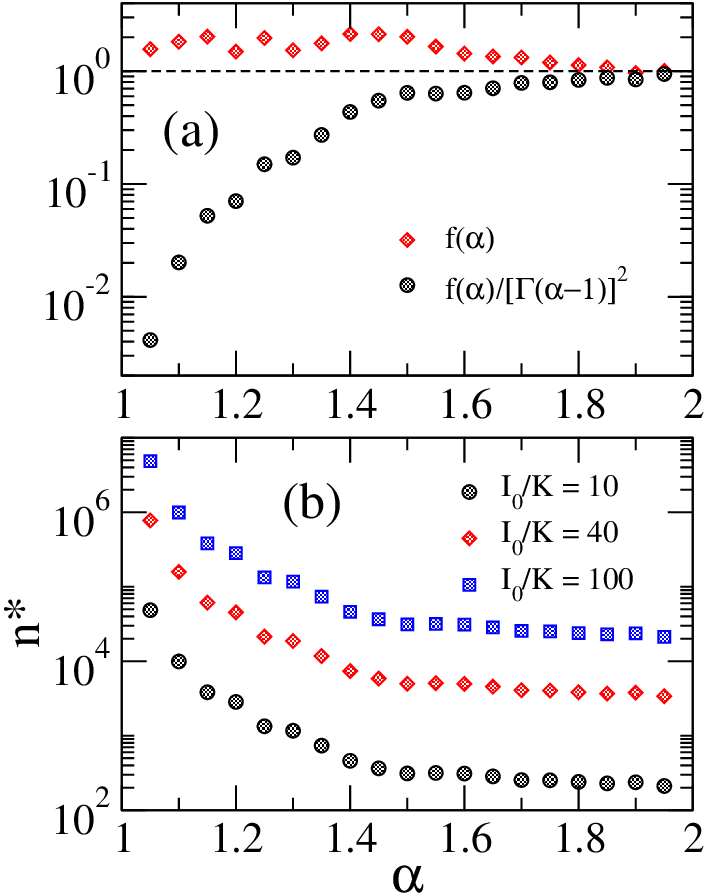}
\caption{
(a) $f(\alpha)$ and $f(\alpha)/[\Gamma(\alpha-1)]^2$.
$f(\alpha)$ is obtained from the power-law fittings of the form $\left< I^2_n \right>_{\tbox{int}}= {\cal C} n^\mu$
to the data of Fig.~\ref{Fig03}(c); i.e.~$f(\alpha)=2{\cal C}[\mu(\alpha) + 1][\Gamma(\alpha-1)]^2/K^2$, here with
$K=10^3$.
(b) $n^*(\alpha)$ for three ratios $I_0/K$; see Eq.~(\ref{nCO}).}
\label{Fig01}
\end{figure}

In analogy with Eq.~(\ref{I2RLfSM}) and taking into account the scaling given in Eq.~(\ref{scalingA}),
we surmise 
\begin{equation}
\label{I2CfSM}
\left< I^2_n \right>_{\tbox{C-fSM}} = I_0^2 + \frac{K^2}{2} 
\frac{f(\alpha)}{[\Gamma(\alpha-1)]^2} n^{\mu(\alpha)} ,
\end{equation}
which leads to
\begin{equation}
\label{I2intCfSM}
\left< I^2_n \right>_{\tbox{int}} = I_0^2 + \frac{K^2}{2} 
\frac{f(\alpha)}{[\Gamma(\alpha-1)]^2[\mu(\alpha)+1]} n^{\mu(\alpha)} .
\end{equation}
Indeed, from the power-law fittings made in Fig.~\ref{Fig03}(c) we can extract $f(\alpha)$, which is
plotted in Fig.~\ref{Fig01}(a).
Notice that $f(\alpha)\sim 2$ for $\alpha<0.5$, while it tends to one for $\alpha\to 2$, as expected. 
In Fig.~\ref{Fig01}(a) we also plot the ratio $f(\alpha)/[\Gamma(\alpha-1)]^2$, which is relevant since 
it appears in Eq.~(\ref{I2CfSM}) and together with the power $\mu(\alpha)$ is one of the key 
differences between this equation and Eq.~(\ref{I2RLfSM}) for the RL-fSM.

In Fig.~\ref{Fig02} we include Eq.~(\ref{I2intCfSM}), as blue-dashed lines and observe
a reasonable good correspondence with the data. We believe that the correspondence between 
Eq.~(\ref{I2intCfSM}) and the data should improve by increasing the number of orbits used in the 
computation of $\left< I^2_n \right>_{\tbox{int}}$. Moreover, we also note an important deviation
of the data from Eq.~(\ref{I2intCfSM}) for very short times, $n<10$, where Eq.~(\ref{I2intCfSM}) 
completely fails.

\section{Discussion and conclusions}

It is relevant to note that Eq.~(\ref{I2CfSM}) can be used to define an effective parameter controlling
the strength of nonlinearity $K_{\tbox{eff}}$ in the C-fSM as
\begin{equation}
\label{I2CfSMKeff}
\left< I^2_n \right>_{\tbox{C-fSM}} = I_0^2 + \frac{K_{\tbox{eff}}^2(\alpha)}{2} n^{\mu(\alpha)} ,
\end{equation}
with
\begin{equation}
\label{Keff}
K_{\tbox{eff}}(\alpha) \equiv \frac{\sqrt{f(\alpha)}}{\Gamma(\alpha-1)} K.
\end{equation}
Indeed, the form of Eq.~(\ref{I2CfSMKeff}) is very convenient because allows a direct comparison 
with Eq.~(\ref{I2RLfSM}) which describes the squared average action of the RL-fSM but also of
the CSM and the DSM.
Thus, it is relevant to stress that, since $K_{\tbox{eff}}(\alpha)\propto 1/\Gamma(\alpha-1)$,
$K_{\tbox{eff}}\to 0$ for $\alpha\to 1$ while $K_{\tbox{eff}}\to K$ for $\alpha\to 2$.

Moreover, from the ratio  
\begin{equation}
\label{I2g0scaling}
\frac{\left< I^2_n \right>_{\tbox{C-fSM}}}{I_0^2} = 1 + \frac{n^{\mu}}{n^*} ,
\end{equation}
we can identify the crossover time
\begin{equation}
\label{nCO}
n^*(I_0,K,\alpha) \equiv 2\frac{I_0^2}{K_{\tbox{eff}}^2} = 2\frac{I_0^2}{K^2} \frac{[\Gamma(\alpha-1)]^2}{f(\alpha)}.
\end{equation} 

Notice also that Eq.~(\ref{I2CfSMKeff}) allow us to define the scaling laws
\begin{equation}
\label{scaling1}
\left< I^2_n \right>_{\tbox{C-fSM}} = \left\{
\begin{array}{ll}
\propto K_{\tbox{eff}}^2 n^\mu , & \ \ \mbox{when} \ \ I_0 \ll K_{\tbox{eff}} , \\
\left.
\begin{array}{ll}
\approx I_0^2 , & n<n^* \\
\propto K_{\tbox{eff}}^2 n^\mu , & n>n^*
\end{array} 
\right\}
& \ \ \mbox{when} \ \ I_0 \gg K_{\tbox{eff}} .
\end{array}
\right.
\end{equation}
Here, $n^*$ separates the regime of constant action and the subdiffusive regime 
when $I_0 \gg K_{\tbox{eff}}$.  
However, note that since $n^*\propto [\Gamma(\alpha-1)]^2$, and $\Gamma(\alpha-1)$ diverges for
$\alpha\to 1$, in practice, the subdiffusive regime may never be approached for $\alpha\to 0$.
As examples, in Fig.~\ref{Fig01}(b) we plot  $n^*(\alpha)$ for three ratios $I_0/K$.
Notice that for $\alpha\sim 1.05$ and $I_0/K=100$, $n^*$ is already of the order of $10^7$.

Finally, it is relevant to recall that subdiffusive dynamics has already been reported for the 
CSM, see e.g.~\cite{MR13,PDCS21,MMS22}. Specifically, $\mu=0.9$~\cite{MR13} and 
$\mu=0.25$~\cite{PDCS21} were found for the CSM with $K=7$ and $K=1.46$, respectively.
However, the anomalous diffusion shown in Refs.~\cite{MR13,PDCS21,MMS22} is produced 
by stickiness around islands of stability in a mixed phase space.
In contrast, the mechanism for the anomalous diffusion we report here is completely different:
Anomalous diffusion in the C-fSM is a consequence of the memory, imposed by the Caputo 
fractional derivative, in the equation for the action.
 
Given that subdiffusion in the C-fSM can continuously be tuned with the parameter $\alpha$ 
(from weak subdifussion, $\mu\sim 1$, to strong subdiffusion, $\mu\sim 0$), the 
C-fSM may serve as a reference model to prove and characterize the effects of subdiffusion 
in other dynamical properties of interest, such as scattering and transport properties.

\section*{Acknowledgements}

J.A.M.-B. thanks support from CONAHCyT-Fronteras (Grant No.~425854) 
and VIEP-BUAP (Grant No.~100405811-VIEP2024), Mexico.


\end{document}